# Intelligent resonance tracking of a microwave plasmonic resonator for compact wireless sensors

Xuanru Zhang,[1, 2, 3] Jia Wen Zhu,[1, 2, 4] and Tie Jun Cui[1, 2, 3] *

1. *State Key Laboratory of Millimeter Waves, Southeast University, Nanjing 210096, China*

2. *Institute of Electromagnetic Space, Southeast University, Nanjing 210096, China*

3. *School of Information Science and Engineering, Southeast University, Nanjing 210096, China*

4. *School of Electronic Science and Engineering, Southeast University, Nanjing 210096, China*

* E-mail: tjcui@seu.edu.cn

**Plasmonic sensing has been in the spotlight for decades, the concept and applications of which have been generalized to spoof surface plasmons (SSPs) in the microwave band. Here, we report a compact and wireless sensor within a printed circuit board size of 18 mm × 12 mm, tracking the resonance frequency shift of a microwave plasmonic resonator via a software-defined scheme. The microwave plasmonic resonator yields a deep-subwavelength size, enhanced sensitivity, and a good electromagnetic compatibility performance. The software-defined resonance tracking scheme minimalizes the hardware circuit and the consumed spectrum resources, and makes the detection intelligently adaptive to the target resonance, with a signal-to-noise ratio of 69 dB and a data rate of 2272 measuring points per second. The sensor has been validated via acetone vapor concentration sensing, while its applications can be widely extended by replacing the transducer materials. This approach provides compact, sensitive, accurate and intelligent solutions for resonant sensors in the Internet of things (IoT).**

Surface plasmons (SPs) have been proved to be an invaluable technology for optical sensing[1], which has been applied to diverse fields including antigen-antibody interactions[2], cancer biomarker detection[3], food safety[4], gas concentration monitoring[5], etc. The subwavelength confinement of light in plasmonic structures enhances light-matter interactions and makes them ultra-sensitive to the surrounding dielectric environment[6-8]. Being in the spotlight for decades, plasmonic sensing become a well-established technology and has been developed for commercial instruments, while there are still ongoing and enticing new breakthroughs[8-10].

The concept of SPs has been generalized to spoof surface plasmons (SSPs) in microwave and terahertz frequencies, based on artificial metallic structures which behave as effective Drude metals while metals are almost perfectly conductive at these lower frequencies[11-14]. Both propagating and locally resonating modes of SSPs have been demonstrated using designed metallic patterns on printed circuit boards (PCBs)[15-17]. Besides resembling the superiorities of optical SPs, SSP sensing exhibits more unique significances in microwave frequencies[18]. From the principal point of view, due to much lower metal loss, SSP resonances present higher quality factors (Q-factors) and multipolar higher-order modes, combining the plasmonic evanescent enhancement and microcavity standing-wave enhancement of sensitivity[18,19]. From the engineering point of view, the compatibility of SSPs to PCBs makes them integratable with subsequent analog and digital circuits and suitable for portable and compact sensors in the Internet of things (IoT)[18,20,21]. Compared with conventional microwave resonators, deep-subwavelength modal confinement of SSP resonators leads to compact sizes, high sensitivity, high immunity to electromagnetic interferences (EMI), and low radiated emission[18]. As a newly arisen field, microwave plasmonic sensing booms rapidly with attempts emerging in liquid concentration sensing[22], monitoring of the engineering structures[23], etc.

For an electromagnetic (EM) resonant sensor, the detection system converting the resonance shifts into voltage signals is crucial, which determines the overall detection limit and the total size of the sensing system. A direct way is scanning the frequency back and forth over the resonance band to find the dip position. It is the most adopted method in both microwave and optical resonant sensing, using either bench top instruments such as vector network analyzers (VNA) for microwaves and spectrometers for optics, or using tuning and detecting circuits in recently developed compact sensors[24]. However, the scanning process is time-consuming, and it consumes too much spectrum resources. For simplification, intensity measurement at a single frequency is often carried out in both instruments[25] and wearable sensors[26], but it sacrifices the accuracy and is susceptible to signal fluctuations. Conventionally, detection schemes involving feedback locking loops can lead to high signal-to-noise ratios (SNR), and work in narrow bandwidths just around the resonances. In the microwave band, phase-locked loops (PLL) are adopted to track the resonance[27,28]. However, the circuit scale is large, and the settings have been fixed in designing which are not versatile for variant sensing scenarios. Pound-Drever-Hall (PDH) locking technique, which was originally proposed for laser frequency stabilization via a feedback control loop[29,30], has been employed in optical microcavity sensing, using the error signal which is proportional to the resonance shift as the sensing response[31-33]. PDH technique has very extensive and versatile applications for its outstanding accuracy and time resolution, including interferometric gravitational-wave detector[34], high-fidelity quantum control,[35] etc. However, manipulation of such a precise locking loop and determination of the control parameters is often tricky, although many investigations have been devoted to automatic locking and relocking methods[36,37].

In this article, we report a compact and wireless sensor, which is composed of a microwave plasmonic resonator (MPR) and its detection and data processing system. The MPR demonstrates a deep-subwavelength resonator size, greatly enhanced sensitivity, and a good electromagnetic compatibility (EMC) performance. A developed PDH locking scheme with automatically determined control parameters is executed in a microcontroller unit (MCU) to track the resonance shift of the MPR. The proposed software-defined resonance tracking scheme minimalizes the circuit hardware and the consumed spectrum resources, and makes the detection intelligently adaptive to the target resonance. The sensor communicates and interacts with a smartphone via Bluetooth. The total size of the sensor PCB is 18 mm×12 mm. An SNR of 69 dB and a data rate of around 2272 measuring points per second has been demonstrated using an 8 MHz 12-bit MCU. Its sensing function is demonstrated by acetone vapor concentration sensing, while its applications can be widely extended using different transducer materials. This scheme provides compact, sensitive, accurate, and intelligent solutions for resonant sensors in the IoT.

# Results

## Overview of the system architecture.

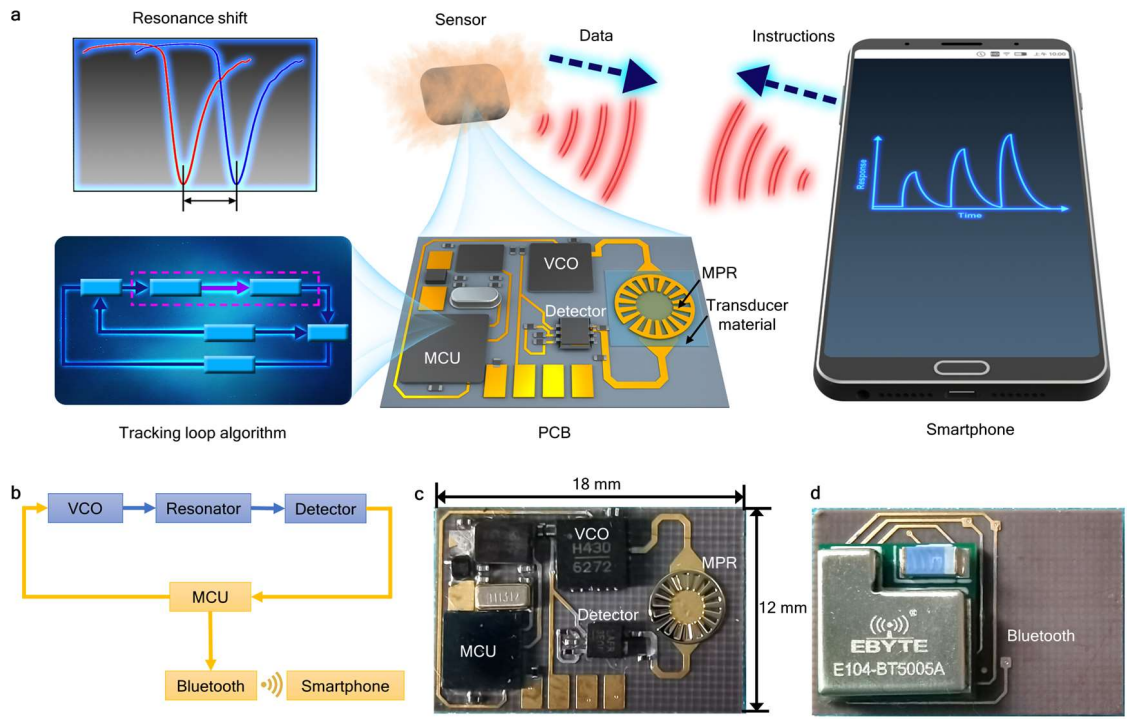

**Fig. 1 | A compact and wireless sensor tracking the resonance shift of an MPR. a**, Schematic of the microwave plasmonic sensor and the sensing network. The sensor transfers data to, and receives instructions from a smartphone. The sensing signal originates from the resonance frequency shift of the MPR. The resonance tracking loop is realized by both the hardware circuit and the algorithm inside the MCU. **b**, Block diagram of the hardware system of the sensor. The blue blocks indicate the main microwave analog circuit, and the yellow blocks indicate the digital computation and communication circuit. **c**, Photograph of the sensor PCB in the top view. **d**, Photograph of the sensor PCB in the bottom view.

The schematic diagram of the compact and wireless microwave plasmonic sensor and its sensing network is shown in Fig. 1a. The sensing signal originates from the resonance frequency of the MPR, which will shift with the effective permittivity of the transducer materials attached to the resonator surface. The detection circuit, which converts the resonance shifts of the MPR to voltage signals, is minimalized and mainly composed of a voltage-controlled oscillator (VCO), a detector diode, and an MCU as shown in Fig. 1b. A Bluetooth module is used to send the measured data to and get instructions from a smartphone. All hardware devices including periphery circuits have been covered in a multilayer PCB with a size of 18 mm 12 mm, except a button cell, as shown in Fig. 1c and d (for details see Methods).

The function of the plasmonic sensor is versatile and depends on the transducer materials attached to the MPR surface (see Methods), which is validated by acetone vapor concentration sensing based on polydimethylsiloxane (PDMS) films in this article. The PDMS film will swell after absorbing the acetone vapor, and its effective permittivity will increase[38,39]. Therefore, the resonance frequency will decrease with increasing acetone vapor concentration. We remark that, this paper emphasizes physical and electric investigations instead of chemical transducers. PDMS films are chosen for their simple accessibility and easy manipulation for the non-specialist in chemistry. The overall vapor concentration sensing performance can be improved by replacing PDMS with advanced transducer materials which can produce more obvious permittivity changes when absorbing the target gas[40-42].

The resonance tracking loop is mainly realized via the algorithm in the MCU. Instructed by the smartphone, the sensor can work under both the frequency scanning mode and the resonance tracking mode. The control parameters for the resonance tracking mode can be automatically calculated according to the transmittance resonance spectrum got from frequency scanning. The output frequency of the VCO is controlled via the digital-to-analog converter (DAC) voltage from the MCU, and the transmitted power of the MPR is detected by the detector diode and sent to the analog-to-digital converter (ADC) of the MCU. Thus, the horizontal and vertical coordinates of the on-PCB measured transmittance spectrum of the MPR will be the DAC and ADC voltages in the following discussions.

The working frequency of the demonstrated sensor is around 4.9 GHz. Although the proposed sensor exhibits a good EMC performance, the working frequency can be conveniently redesigned to other microwave frequencies according to the specific EM environment to avoid any possible EMI. The working frequency is around the resonance and can be tuned by the size and the pattern of the MPR. The VCO and detector should be chosen covering the working frequency. The transducer materials usually work for a broad spectrum ranging from microwaves to optics.

**Design and analysis of the MPR.**

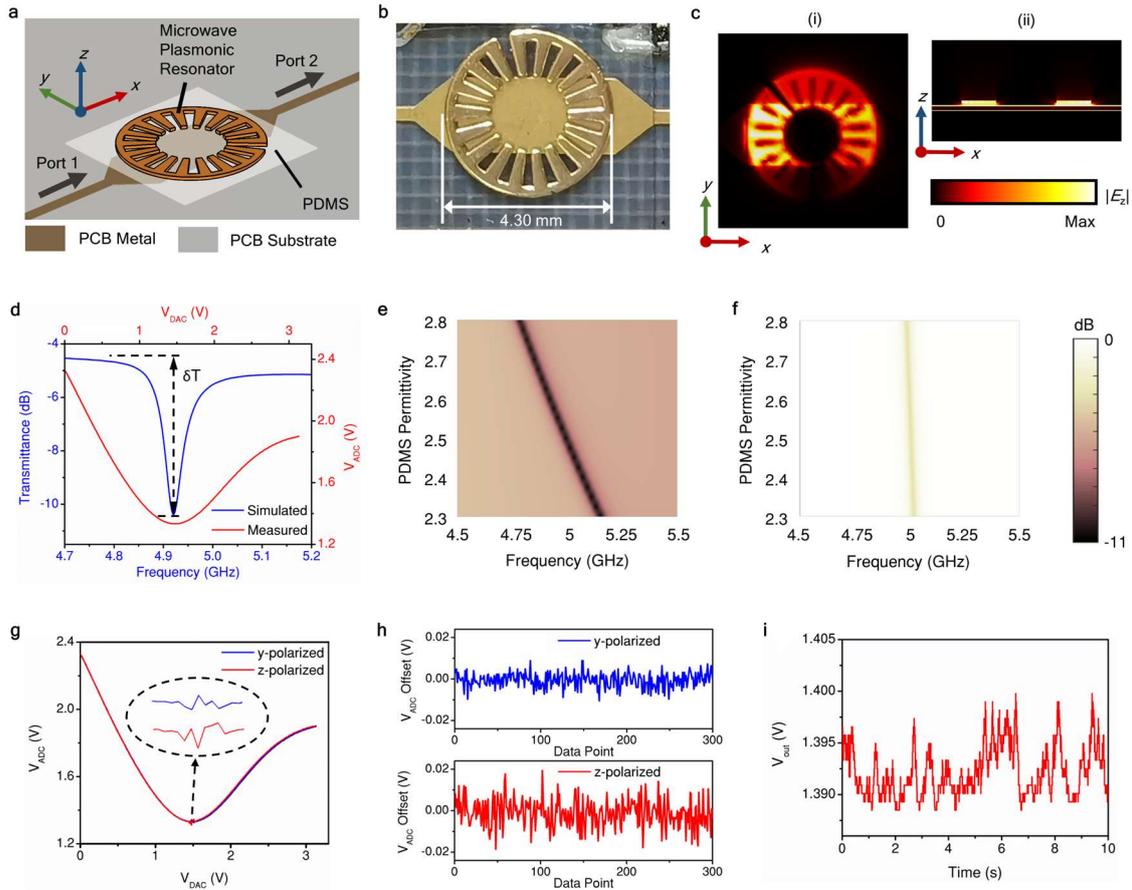

**Fig. 2| Design and measurement of the MPR. a**, Schematic of the MPR and its excitation structure. **b**, Detail photograph of the fabricated MPR, attached to the exciting microstrip pattern with an interlayer of transparent PDMS film. **c**, Simulated $E_z$ distributions at the resonator surface (i) and the cross-section in the *xz*-plane (ii). **d**, Transmittance spectra of the MPR obtained via electromagnetic simulation and on-PCB measurement respectively. **e**, Mapping of the MPR transmittance spectrum varying with the relative permittivity of PDMS. **f**, Mapping of the transmittance spectrum varying with the relative permittivity of PDMS, for the contrastive MRR. (e) and (f) share the same color bar. **g,** Measured transmittance spectra under CW interfering radiations of 4.94 GHz in *y*- and *z*- polarizations. **h**, Offsets of the ADC voltage at the bottom of the resonance dip, measured under CW interfering radiations of 25 dBm and at 4.94 GHz,

in *y*- and *z*- polarizations respectively. **i**, Noise level of the resonance tracking detection, measured under CW interfering radiations of 25 dBm and at 4.94 GHz in *z*- polarization.

The proposed MPR and its excitation structure are demonstrated in the schematic in Fig. 2a, and in the photograph in Fig. 2b. A standalone MPR of 0.1 mm thick (which is made of stainless steel and plated by copper, for details see Methods) hangs over the PCB, with an intermediate PDMS film of 50 μm thickness (for the detailed layout see Supplementary Fig. 1 and 2). The field distributions of the MPR resonance mode are demonstrated in Fig. 2c. The plasmonic resonance mode is strongly compacted by the slit which breaks the spatial symmetry (for detailed mode analysis see Supplementary Note 1). Sandwiching the PDMS film between the MPR and the top microstrip pattern of the PCB, makes the EM fields concentrate in and interact adequately with the PDMS. Therefore, the sensitivity is greatly enhanced compared with common geometries which print the resonators on the microstrip layer (top metal layer) and lay the transducer materials on top of the resonator. Moreover, the resonance intensity δT, which is defined as the difference between the top and the bottom of the transmittance resonance curve (Fig. 2d), is greatly enhanced by this sandwiching structure. The resonance intensity is one of the key issues of on-PCB measurement, which means full use of the specific ADC resolution and affects the overall noise level (see Supplementary Note 3). Although capacitive coupling is commonly used to enhance the resonance intensity of deep-subwavelength resonators, it greatly decreases the sensitivity as the EM fields are partly attracted into the high-permittivity but inert capacitance[19,43]. The proposed sandwiching structure enhances the sensitivity and resonance intensity simultaneously.

For an intuitional sense of its superiorities, the proposed MPR is quantitatively compared with a microstrip ring resonator (MRR) resonating at the same band (for details of the MRR see Supplementary Note 2). The proposed standalone MPR with the sandwiching PDMS structure is also compared with the same MPR pattern printed on PCB in Supplementary Note 1. The On-PCB measured and simulated transmittance spectra of the proposed MPR are presented in Fig. 2d (for details of the EM simulation see Methods and Supplementary Note 1). The *x*-axes have been calibrated using the frequency tuning curve of the VCO. The simulated resonance intensity is 0.44 on a linear scale (which means 44% power from the microstrip circuit is fed into the resonator). This resonance intensity is almost double-fold of the case that the same MPR is printed on the PCB (see Supplementary Table 1 in Supplementary Note 2). The MPR resonates at around 4.92 GHz (Fig. 2d), which means the resonance is confined in diameter of 1/14.2 wavelength. Due to the deep-subwavelength confinement, the local electric fields are strongly enhanced, which indicates enhanced local sensitivity. The diameter of the contrastive MRR is 3.1-fold that of the MPR, and, the maximum electric field enhancement of the MPR is 2.7-fold that of the contrastive MRR. As demonstrated in Fig. 2e and f, the resonance shift sensitivity and the resonance intensity of the proposed MPR are obviously enhanced than the contrastive MRR. Local sensitivity $S_L$ is defined to evaluate the sensing response within unit sensor size as:

$$S_L = df_r/d\varepsilon \cdot (1/S) \quad (1)$$

in which $f_r$ is the resonance frequency, $\varepsilon$ is the relative permittivity of the PDMS film, and $S$ is the area of the resonator. $S_L$ is calculated via EM simulation as presented in Supplementary Note 2. The local sensitivity of the proposed MPR is 108-fold of the contrastive MRR resonating at the same band, and is 9.7-fold of the case that the same MPR pattern printed on the PCB top layer.

Comparing the two transmittance spectrum curves in Fig. 2d, there is an obvious degradation of the on-PCB measured Q-factor. Besides the fabrication imperfections, the phase noise i.e., broadened linewidth of the VCO, varying VCO output power depending on the tuning voltage, nonlinearity in VCO output frequency tunability, nonlinear response of the detector to the input power, and the frequency-dependent response of the detector all participate in the distortion of the on-PCB measured transmittance resonance spectrum (for detailed analysis see Supplementary Note 4). Calibrating the DAC/ADC voltages to VCO frequency and transmittance power, the on-PCB measured Q-factor is 18.5. Although the on-PCB measured Q-factor degrades greatly compared with the simulated value of 103.2, it is still a relatively high value compared with other conventional deep-subwavelength resonators[19].

It is difficult to couple free-space waves into deep-subwavelength resonators, therefore, high immunity to EMI can be expected. The EMI immunity test is carried out under a continuous wave (CW) radiation of 25 dBm from a horn antenna located at a distance of 100 mm (for details see Methods and Supplement Fig. 12). Varying the interfering radiation frequency, it is found that the frequency at resonance (which is measured to be 4.94 GHz and slightly deviates from the simulated value of 4.92 GHz) causes the strongest interferences. As shown in Fig. 2g, the interfering radiation causes slight interferences to the transmittance curves. The offsets of the ADC voltage just at resonance, which are measured by fixing the VCO output frequency (i.e., the DAC voltage) and collecting the detector response (i.e., the ADC voltage), are shown in Fig. 2h. Larger offsets are generated by the *z*-polarized interference as it is the main polarization of the resonance mode and couples more with the MPR. The standard deviation is 0.0067 V in Fig. 2h for the *z*-polarized interference, which corresponds to an SNR of 46 dB referring to an averaged $V_{ADC}$ of 1.33 V. During measurement, not only the resonator but also the whole sensor circuit will couple with the interfering radiation. As the whole sensor PCB is also in the subwavelength scale (the PCB size is 18 mm × 12 mm and the incident wavelength is 60.7 mm), the interfering power coupled from other non-resonating parts of the PCB is weaker. Besides the frequency scanning mode, the sensor is also measured under resonance tracking mode. The total SNR of the resonance tracking measurement ($V_{out}$ is defined and discussed in the next section) degrades from 69 dB to 55 dB (Fig. 2i), under interfering radiation of 25 dBm in *z*-polarization at 4.94 GHz. Moreover, from the emission point of view, the radiative efficiency of the deep-subwavelength MPR is -17 dB. The sensing resonator causes little radiated emission, and the radiated emission of the whole sensor mainly comes from the Bluetooth module.

**Intelligent resonance tracking detection.**

The resonance shift of the MPR is detected by a developed PDH scheme, operating discretely in the self-programmed MCU algorithm with intelligently determined control parameters. The block diagram is shown in Fig. 3a, and the conceptual basis is shown in Fig. 3b. A square-wave modulation signal is used, which is implemented via hopping around the resonance up and down and is equivalent to only two sampling points per period for a sinusoidal modulation signal. Therefore, the square-wave signal leads higher data rate than other modulation functions based on the same clock configuration of the MCU. The resonance tracking effects are the same except for a constant coefficient in the error signal formula, as presented via mathematic derivations and Simulink simulations in Supplementary Note 5 and 6. Due to the discrete operation way in our MCU algorithm, the effective modulation rate and the data rate is determined considering both the rising time of the modulation voltage and the time delay in the operation loop (see Supplementary Note 7 and 8). A data rate of 2272 measuring points per second is stably achieved, based on an 8 MHz 12 Bit MCU and a baud rate of 921600 bps.

The parameters for PDH locking are calculated automatically according to the transmittance spectrum, which has been recorded from frequency scanning (green curve in Fig. 3b). Starting $V_0$ in Fig. 3a (which is equivalent to $f_0$ in Fig. 3b) is determined from the resonance dip. The amplitude of the modulation signal ($A_m$) should be generally proportional to the resonance bandwidth, or the response transmittance differential generated in one modulation process may be overwhelmed by noises. As shown in Fig. 3b, T' is calculated from the T spectrum, and $A_m$ is determined to be 1/8 of the width between the maximum and minimum of T'. (To present a neat figure, only one T' curve is shown in Fig. 3b, which is the one resonating at $f_1$. The calculation of

A$_m$ is demonstrated on this blue T' curve, while it is calculated from the transmittance curve resonating at the original position of $f_0$ in the real operation.) The coefficient of 1/8 is experiential, as we test resonances of different Q-factors and find 1/8 leads to good SNRs in most cases. Calculating from the on-PCB measured transmittance spectrum in Fig. 2b and calibrating the horizontal coordinate into frequency, A$_m$ equals 6 MHz. During resonance tracking, the instant working bandwidth is 12 MHz under the working frequency of 4.9 GHz. The consumed spectrum resource is much suppressed compared with the frequency scanning method.

The three control parameters $K_p$, $K_i$, and $K_d$ of the proportional-integral-derivative (PID) controller are crucial for the stable operation and accurate detection of the PDH locking loop. Determination of these parameters is commonly empirical and tricky, although there are tuning guides such as the Ziegler-Nichols method[44]. The PID controller produces the feedback voltage to V$_{DAC}$ based on the error signal (taken as T' in the MCU algorithm). In the discrete implementation, the integral term evolves into a superposition of every feedback step, while $K_p$ and $K_d$ are set to be zero. For each step (the shift from the green curve to blue curve in Fig. 3b), the feedback is defined to be equal to the frequency shift, so that the VCO frequency can be shifted to the instantaneous new resonance position in one single step, in a linear approximation of T'. As interpreted from the modulation triangle in Fig. 3b, the feedback signal in one step is proportional to T' and equals:

$$-\Delta f = K_i \cdot T', \tag{2}$$

in which the automatically calculated $K_i$ value ($K_i$=K) is determined to be reciprocal of the slope of T' as:

$$K_i=K=1/T''. \qquad (3)$$

T'' is calculated using a linear least square regression of T' within [-3$A_m$, 3$A_m$] interval.

Reducing $K_i$ from K, the response will gradually approach the target signal in several aggregations (which can be interpreted from Equation (2)). For our sensor where the data rate is high enough compared with the target signal varying rate, smaller $K_i$ values are preferred for lower noise levels, as large $K_i$ may lead to overshoots and oscillations in the PID control. (The effects of $K_i$ in PDH resonance tracking are analyzed via Simulink simulations in Supplementary Note 6 and 9, for both the ideal noiseless cases and the cases with thermal noises.) As shown in Fig. 3c, the measured SNRs are 48 dB, 53 dB, 60 dB, and 62 dB for $K_i$=K, 0.5K, 0.2K, and 0.1K. However, if an excessively small $K_i$ value is adopted (such as $K_i$=0.001K), the response $V_{out}$ signal may lag the target signal. The automatically calculated parameters make the PDH locking scheme intelligently adaptive to the sensing situations.

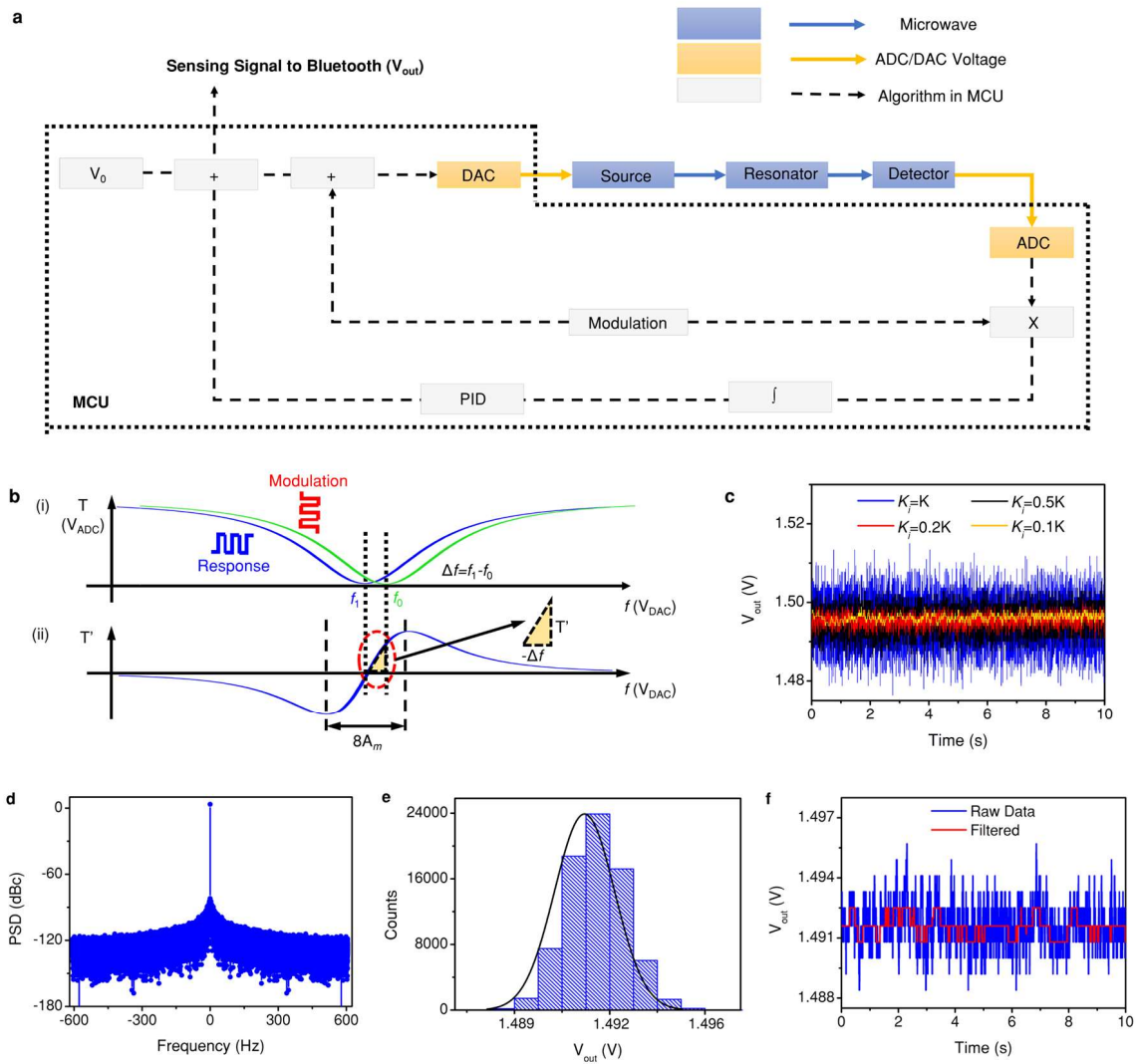

**Fig. 3| Intelligent resonance tracking scheme. a,** Block diagram of the resonance tracking loop. The signals in the hardware circuit are denoted by solid lines while the signals in the algorithm are denoted by dashed lines. The blue color indicates the microwave signals, while the yellow color indicates the ADC/DAC voltages. The blocks within the dotted black line represent both the hardware and the algorithm in the MCU. **b,** Conceptual basis of the resonance tracking scheme using a square-wave modulation signal, and the automatic calculation method of $A_m$ and K. **c,** Noise level of the resonance tracking detection, when $K_i$ is set to be K, 0.5 K, 0.2 K, and 0.1K. **d,** PSD of the raw data of the noise level when $K_i$=0.1K. **e,** Histogram and fitted Gaussian

distribution function of the raw data of the noise level when $K_i$=0.1K. **f,** Raw data and the filtered data of the noise level, when $K_i$ is set to be 0.1K. The window size of the median filter is 300.

The power spectrum density (PSD), which describes the power distributions over frequency and is defined as the Fourier transform of the $V_{out}$(t) noise level is presented in Fig. 3d. The signal's histogram with the fitted Gaussian distribution function are shown in Fig. 3e. ($K_i$=0.1K in Fig. 3d and e.) The noise is of Gaussian type, however, it's not white as the noise level is higher at lower frequencies. (The thermal noise of the hardware circuits is measured at $V_{ADC}$ and equals 67 dB. The thermal noise is a Gaussian white noise. The Gaussian but not white noise characteristic of $V_{out}$ is verified via Simulink. For details see Supplementary Note 9.) The noise characteristic is consistent with those reported in PDH locking of optical microcavities[31,32]. In the software-defined PDH scheme, the noise level can be conveniently processed by a median filter. As shown in Fig. 3f, the raw data and data processed with a median filter with a window size of 300 are presented. The SNR can be further improved from 62 dB to 69 dB by the median filter ($K_i$=0.1K).

**Function demonstration by acetone vapor concentration sensing.**

As a demonstration example, the compact and wireless sensor is validated by acetone vapor concentration sensing using PDMS films as the transducer materials. The operation flow of the sensor is presented in Figure 4a and Supplementary Video 1. The curves are dynamically displayed in both the frequency scanning and the resonance tracking modes. PDH locking starts with automatically calculated $A_m$ and $K_i$=K. The

PID parameters can be manually modulated in the PID setting page, with automatically calculated values displayed for reference. The displayed resonance tracking data has been processed by a median filter with a window size of 300. Using the "smooth" function, the displayed data will be filtered by a median filter with a window size of 3000. The "Relock" function on the resonance tracking page is designed to restart the resonance tracking process with automatically calculated control parameters. This is designed for a common issue in PDH locking that losing locking sometimes happens when the target signal varies abruptly or the PID parameters are not suitable.

Fig. 4b shows the photograph of the sensor working with a smartphone. Only a button cell is needed for the operation, besides the PCB. Long wiring is used for a clear demonstration, while the whole package could be further compacted. For acetone vapor sensing, a dynamic gas mixing system is constructed to control the acetone vapor concentration (see Methods and Supplementary Fig. 22). The interflow rate is fixed at 1L/min to lower possible temperature jitters caused by the flow. Realtime responses of the sensor to increasing acetone concentration and sequential acetone pulses of increasing concentrations are presented in Fig. 4c and d. The response is reproducible and the acetone absorption is reversible. Qualitative monitoring of acetone evaporation is also demonstrated in Supplementary Video 2. We remark again that, this measurement is only a concept demonstration. Replacing commercially available PDMS films with transducer materials which produces more significant permittivity changes, will directly lead to more advantageous sensitivity and detection limits, and will fully fulfill the sensitivity advantages of the MPR..

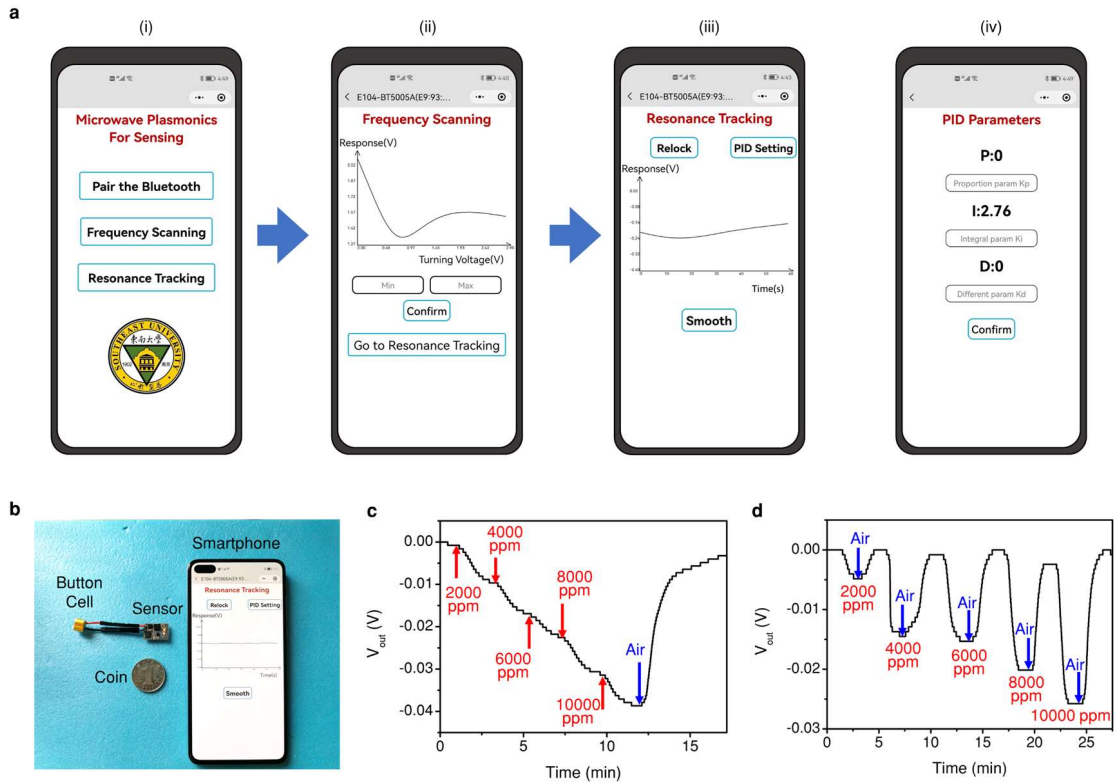

**Fig. 4| Acetone vapor concentration sensing experiments. a,** The smartphone user interfaces for the function menu (i), frequency scanning (ii), resonance tracking (iii), and user-defined PID parameters setting (iv). **b,** Photograph of the sensor working with a smartphone. The size of the sensor is compared with a 1 Chinese Yuan coin, the diameter of which is 25 mm. **c,** Measured response of the sensor to increasing concentrations of acetone vapor. **d,** Measured response of the sensor to acetone vapor pulses of different concentrations.

## Conclusions

We have reported a compact and wireless microwave plasmonic sensor, and demonstrated its application via acetone vapor concentration sensing. We show that, the advantages of the MPR and the advantages of the software-defined resonance tracking scheme are combined, realizing a compact PCB size of 12 mm×18 mm, a high SNR of 69 dB, a high data rate of 2272 measuring data points per second, enhanced sensitivity, a good EMC performance, and intelligently adaptive running of the sensor. A 108-fold local sensitivity enhancement compared with conventional MRR has been demonstrated. The high EMI immunity and low radiative emission of the MPR, combined with the narrow spectrum bandwidth taken by the resonance tracking detection method, endow the proposed sensor with a good EMC performance. This work envisions the broad prospect of microwave plasmonic resonators in compact and wireless sensors for the IoT, and further supplies a compact, intelligently adaptive, and versatile resonance tracking solution for a vast variety of electromagnetic resonant sensors from optical to microwave frequencies, and even for mechanical resonant sensors.

## Methods

**Transducer materials and corresponding sensing functions.** For electromagnetic resonant sensors from microwave and optical ones, their functions are versatile and rely on the transducer materials attached to the resonator surface. The fundamental physical sensing signal originates from the effective permittivity changes of the transducer materials. The transducer materials should be dielectric, similar to those in capacitive sensing in lower frequencies. Semiconductors with larger loss tangents will broaden the resonances and degrades the sensing performance.

The proposed sensing system can also be used for other gas sensing replacing the PDMS films with other transducer materials, such as polyethyleneimine for formaldehyde sensing[45], $TiO_2$ for ammonia sensing[46], etc. The sensor can also work in an aqueous medium for label-free biomedical sensing, replacing the PDMS film with a microfluidic channel. Printed the MPR and the circuits on a flexible substrate, this design could also work for strain sensing[26]. Of course, the details of the MPR and the circuit should be designed according to each specific case.

**Materials.** The PCB substrates of 0.127 mm and 0.508 mm thick are chosen to be TLY-5 series with a relative permittivity of 2.2 and a loss tangential of 0.0009, which are obtained from the Taconic Advanced Materials Co., Ltd. The prepreg is chosen to be Rogers RO4450F. The PDMS films of 50 μm thick were purchased from the Hangzhou Bald Advanced Materials Co., Ltd.

**Design and fabrication of the Sensor circuit.** The sensor circuit comprises a voltage regulator, a VCO, an MPR, a detector diode, an MCU, a Bluetooth module, and their periphery circuits. These devices are assembled on a piece of multilayer PCB.

*Fabrication of the multilayer PCB.* The sensor PCB is fabricated using the commercial printed circuit board technique. As shown in Supplementary Fig. 1, the PCB is composed of three copper pattern layers (top, middle, and bottom layers) of 0.035 mm thick, separated by two dielectric substrate layers. The top dielectric substrate is 0.127 mm thick, and the bottom substrate is of 0.508 mm thick. A prepreg layer locates between the middle copper layer and the bottom dielectric substrate. The total thickness of the PCB is 0.737 mm. The top copper layer constitutes the main functional circuit as shown in Fig. 1c, while the bottom copper layer is for the mounting of the Bluetooth module. The main part of the middle copper layer is the ground plane, besides some via holes and wiring for control voltages. The ground plane on the middle copper layer is conducted to the ground pads on the top and the bottom copper layers.

*Design of the MPR.* The EM resonance modes of the MPR are designed and analyzed based on EM simulations in the CST microwave studio. Simulations of the S-parameters are carried out using the Time Domain Solver of the CST Microwave Studio, with an accuracy of -60 dB and a maximum solver duration of 1000 pulses. The accuracy and maximum solver duration settings are crucial for simulations of high-Q resonators. Open (add space) boundaries are imposed, and two waveguide ports are used for simulations of the S-parameters. The dispersion curve is calculated using the Eigenmode Solver of the CST Microwave Studio. Periodic boundaries are imposed on the *x*-direction. The mode analysis is thoroughly discussed in Supplementary Note 1 and 2.

*Fabrication of the MPR.* The standalone MPR is made of stainless steel to guarantee its mechanical strength, via the stamping process using a laser-cutting die. Then it is plated by copper for high surface conductivity. The diameter of the MPR is 4.3 mm with a fabrication tolerance of 0.1 mm, and the thickness is 0.1 mm with fabrication tolerance of 0.02 mm.

*Devices.* The producers and part numbers of the devices are as follows: (1) VCO: Analog Devices, HMC430LP4; (2) Diode Detector: Linear Technology, LTC 5532; (3) MCU: STMicroelectronics, STM32F103RCY6 in WLCSP64 package; (4) Voltage

Regular: Texas Instrument, TLV75530PDQNR; (5) Bluetooth module: Chengdu Ebyte Electronic Technology Co., Ltd., E104-BT5005A; (6) Button cell (rechargeable lithium polymer battery): Shenzhen Yuhuida Electronics Co., Ltd, 601015, 3.7 V 30 mAh. The devices are mounted to the PCB using a soldering machine. The sensor PCB is connected to a button cell via a pair of Dupont wires.

*Power Consumption.* The power consumption is calculated according to the datasheet of the active devices. (1) VCO: 3.3V, 27 mA, 89.1 mW; (2) Diode Detector: 3.3 V, 500 µA, 1.6 mW; (3) MCU: 3.3 V, 5.8 mA, 19.1 mW; (4) Bluetooth module: 3.3 V, transmitting (TX) current, 7.2 mA, receiving (RX) current, 6.4 mA; 23.8 mW (TX mode). Thus, the total power consumption is 133.6 mW when the Bluetooth module working at the TX mode, and is 130.9 mW when the Bluetooth module working at the RX mode.

**Data collection.** The curves in the screenshot of the smartphone user interface and in the Supplementary Videos are recorded via the WeChat Mini program on the smartphone. Data in the curve figures are recorded via a computer using a USB to TTL adapter, for the convenience of data collecting and processing. The measuring data rate of 2272 measuring data points per second is measured when the baud rate of the MCU is set to be 921600 bps, which matches to the highest baud rate of the Bluetooth module.

**Data transferring to and plotting in the WeChat Mini program.** Due to limited pixels on the smartphone, the sensor's data rate of resonance tracking mode is surplus. Therefore, the baud rates of both the MCU and the Bluetooth module are set to be 115200 bps when the sensor is working with a smartphone, and the measuring data rate is lowered to around 1200 measuring points per second. A character string "0000" is sent and a delay of 100 ms is manually added when every 300 data points are sent, for data structure alignment in the WeChat Mini program.

**EMI immunity Test.** The EMI immunity test is carried out in a microwave anechoic chamber. A horn antenna (XB-HA187-15, Beijing Xibao Electronic Technology Co., Ltd) is used as the source, which is fed by 25 dBm power from a signal generator (Keysight N9040B). The main polarization of the radiation is along the short side of the horn. The sensor is attached to a piece of foam via adhesive tapes. The horn antenna is fixed during measurement, and the foam is placed in different orientations to change the polarization relative to the MPR. The distance between the sensor and the horn antenna is fixed at 100 mm during the test, and the sensor is always located aligning with the center of the horn. It can be evaluated from the EM simulation that, the electric field at the sensor position is 123.7 V/m when 25 dBm power is fed to the horn antenna. The photograph of the EMI immunity setup is shown in Supplementary Fig. 12.

**Dynamic gas mixing.** A homemade gas mixing system is used for acetone vapor concentration control. Compressed air and acetone vapor of 10000 ppm come from gas bottles and go through two rotor flow meters (flow range: 0.0-1.0 L/min) respectively and converge at a Tee-junction. The mixed gas goes through a silica chamber with an inlet and an outlet tubing. The inside volume of the silica chamber is 80 mm*80 mm* 30mm. The sensor locates in the chamber during the test. There is a movable ceramic base for the silica chamber, allowing the entrance and exitance of the sensor. The silica chamber and the ceramic base are sealed via air-tight seal strips. The total flow rate is fixed at 1.0 L/min, and the acetone concentration is tuned by the ratio of the two rotor flow meters. The setup schematic and the photograph are shown in Supplementary Fig. 22.

## Acknowledgements

This work was supported by the National Science Foundation of China under Grant Nos. 61701108, 61631007, 61871127, 61890544, 61801117, 61735010, 61731010, 61722106, and 61701107, the National Key Research and Development Program of China under Grant Nos. 2017YFA0700201, 2017YFA0700202, and 2017YFA0700203, the 111 Project under Grant No. 111-2-05, and the Fundamental Research Funds for the Central Universities under Grant No. 2242021k30047.

## Author contributions

T. J. C. and X. Z. conceived the idea and wrote the manuscript. T. J. C. suggested and supervised the work. X. Z. conducted the systematic designs, simulations, and data analysis. J. W. Z. conducted the MCU and the WeChat Mini program coding. X. Z and J. W. Z. carried out the experiments. All authors discussed the results and commented on the manuscript.

## Competing interests

The authors declare no competing interests.

## Additional information

Supplementary Information